\documentclass{PoS}
\usepackage{amssymb,latexsym}
\usepackage{amsmath,amsbsy}
\usepackage{epsfig,bm}
\usepackage{graphicx}
\usepackage{comment}
\DeclareFontFamily{OT1}{pzc}{}
\DeclareFontShape{OT1}{pzc}{m}{it}%
             {<-> s * [1.40] pzcmi7t}{}
\DeclareMathAlphabet{\mathscr}{OT1}{pzc}%
                                 {m}{it}

\def\d{{\delta}}
\def\D{{\Delta}}

\def\L{{\Lambda}}

\def\S{{\Sigma}}

\def\CPT{{$\chi$PT}}

\def\cF{{\mathcal F}}

\def\cL{{\mathcal L}}

\def\cI{{\mathcal{I}}}
\def\cJ{{\mathcal{J}}}

\def\eqref#1{{(\ref{#1})}}
\title{Axial and electromagnetic observables of hyperons in 2-flavor chiral perturbation theory}

\ShortTitle{Hyperon properties in 2-flavor $\chi$PT}

\author{\speaker{F.-J. Jiang}\\
        Department of Physics, National Taiwan Normal University, 88, Sec.4, Ting-Chou Rd., Taipei 116, Taiwan\\
        \email{fjjiang@ntnu.edu.tw}}

\author{B.~C.~Tiburzi\\
        Center for Theoretical Physics, Massachusetts Institute of Technology,
77 Massachusetts Ave, Cambridge, MA 02139, USA\\
\email{bctiburz@mit.edu} }

\author{
A. Walker-Loud\\
Lawrence-Berkeley National Laboratory, One Cyclotron Road, MS 70R0319, Berkeley, CA 94720-8169, USA\\
\email{awalker-loud@lbl.gov}
}

\abstract{Two-flavor chiral expansions provide a useful perturbative framework to study hadron properties. 
Such expansions should exhibit marked improvement over the conventional three-flavor chiral expansion. 
Although in principle one can formulate two-flavor theories for the various hyperon multiplets, 
the nearness of kaon thresholds can seriously undermine the effectiveness of such two-flavor theories in practice. 
We investigate the importance of virtual kaon thresholds on hyperon properties, 
specifically their isovector axial charges and electromagnetic observables. 
In particular 
we uncover the underlying expansion parameter governing the description of virtual kaon thresholds. 
For spin-half hyperons, 
this expansion parameter is under theoretical control. 
As a result, the
virtual kaon contributions are well described in the two-flavor theory by terms analytic in the pion mass-squared. 
For spin three-half hyperons,
however, 
one is closer to the kaon production threshold, 
and the expansion parameter is not as small. 
Breakdown of 
$SU(2)$
chiral perturbation theory is shown to arise from a pole in the expansion parameter associated with the kaon threshold.
We find that, despite the fact that higher-order corrections to the expansion parameter 
is necessary to ascertain whether the two-flavor theory of spin three-half hyperons
remains perturbative, 
there is a useful perturbative expansion for isovector axial charges and magnetic moments of both spin-half and spin 
three-half hyperons. }

\FullConference{The XXVIII International Symposium on Lattice Field Theory, Lattice2010\\
		June 14-19, 2010\\
		Villasimius, Italy}

\begin{document}
\section{Introduction}

An effective description of low-energy QCD is possible provided a 
systematic power counting can be established to order the infinite terms in the Lagrangian of chiral perturbation theory (\CPT) \cite{Gas84,Gas85}. 
An 
$SU(2)$ 
chiral expansion is better suited for this task compared to 
$SU(3)$, 
because the eta mass squared, 
$m_\eta^2$, 
is not particularly small compared to the square of the chiral symmetry breaking scale, 
$\L_\chi^2$. 
The inclusion of baryons can be done systematically by treating the baryon mass, 
$M_B$,  
as a large parameter \cite{Jen91,Jen91_1}.
An 
$SU(2)$ 
expansion for baryons is expected to be more effective than 
$SU(3)$, 
because the latter expansion contains terms that scale linearly with  
$m_\eta / M_B \sim 0.5$. 
The 
$SU(2)$ 
theory  of hyperons exploits the hierarchy of scales
$m_u, m_d \ll m_s \sim \L_{QCD}$. 
Consequently the strange quark mass dependence  
is either absorbed into the leading low-energy constants of 
$SU(2)$, 
or arises through power-law suppressed terms, 
$\sim ( m / m_s)^n$,
which are absorbed into low-energy constants of pion-mass dependent operators.  
Here 
$m$ 
is used to denote the average of the up and down quark masses. 
The resulting theory sums all potentially large strange quark mass contributions to all orders. 
Efficacy of the two-flavor theory strongly depends on the underlying 
$SU(3)$ dynamics. Kinematically, 
hyperons are forbidden to produce kaons through strong decays.
The nearness of strangeness-changing thresholds, 
however, 
can lead to significant non-analytic quark mass dependence
in hyperon observables.
Such dependence may not be adequately captured in the two-flavor theory 
because explicit kaons are absent. 
Due to the size of hyperon mass splittings, 
spin three-half hyperon resonances are particularly sensitive to kaon contributions. A detailed
analysis for the hyperon observables demonstrates 
that the 
$SU(2)$
chiral expansion of kaon loop contributions are 
under control for hyperon masses and isovector axial charges \cite{Tib08,Tib09,Jia09,Jia09_1}. Further, 
for the spin one-half hyperon electromagnetic properties, 
kaon loop contributions are well captured by terms 
analytic in the pion mass squared \cite{Jia10}. 
The same remains true for magnetic moments of the 
spin three-half hyperons. 
Electromagnetic radii and quadrupole moments of the hyperon resonances are shown to be
quite sensitive to the nearby kaon thresholds. 
The 
$SU(2)$ 
expansion of these kaon contributions appears to converge at the physical pion mass, 
however, 
the efficacy of the two-flavor theory does not extend considerably 
far beyond the physical point. Despite the fact that higher-order corrections to the expansion parameter 
is necessary to ascertain whether the two-flavor theory of spin three-half hyperons
remains perturbative, 
there is a useful perturbative expansion for isovector axial charges and magnetic moments of both spin-half and spin 
three-half hyperons. 

\section{Two-Flavor Chiral Expansion}

As a schematic example, 
we consider the mass of the 
$\S$
baryon. 
In 
$SU(3)$
chiral perturbation theory, 
there are contributions from pion,
kaon, 
and eta loops 
at next-to-leading order. 
To this order, the contributions can be written in the form 
$M_\Sigma 
= 
M^{SU(3)}
+ 
a_\pi \,
m_\pi^2 
+
a_K \,
m_K^2
+ 
b_\pi \, 
m_\pi^3
+ 
b_K 
\, m_K^3
+ 
b_\eta \,
 m_\eta^3
$.
The parameter 
$M^{SU(3)}$ 
is the average octet baryon mass in the SU(3) chiral limit, 
while the 
$a_\phi$ and $b_\phi$ 
coefficients depend on the low-energy constants of the heavy baryon theory. 

By virtue of the Gell-Mann--Oakes--Renner (GMOR) relation, 
we can write the kaon mass in the form
$m_K^2 = \frac{1}{2} m_\pi^2 + \frac{1}{2} m_{\eta_s}^2$,
where $m_{\eta_s}$ is the mass of the quark basis $\overline{s}s$ meson. 
Using leading-order chiral perturbation theory and the mass of the neutral pion and the averaged mass-squared of the kaons, 
we have 
$m_{\eta_s}  = 0.688 \, \texttt{GeV}$.
A natural expansion suggests itself:
expand in powers of
\begin{equation} \label{eq:size}
\varepsilon_{SU(2)} 
=
\frac{m_\pi^2 }{ m_{\eta_s}^2 } \sim 0.04
.\end{equation} 
The eta-mass contributions can also be treated using an expansion 
$\varepsilon_{SU(2)}$ through the Gell-Mann--Okubo formula
$m_\eta^2 
= 
\frac{1}{3} m_\pi^2 
+
\frac{2}{3} m_{\eta_s}^2 
$.
Carrying out the expansion in powers of 
$\varepsilon_{SU(2)}$
on 
$M_\Sigma 
= 
M^{SU(3)}
+ 
a_\pi \,
m_\pi^2 
+
a_K \,
m_K^2
+ 
b_\pi \, 
m_\pi^3
+ 
b_K 
\, m_K^3
+ 
b_\eta \,
 m_\eta^3
$, 
we arrive at an expression for the 
$\S$ 
mass perturbed about the 
$SU(2)$ 
chiral limit, 
\begin{equation} \label{eq:MX2}
M_\Sigma 
= 
M_\Sigma^{SU(2)} + a_\pi^{SU(2)} m_\pi^2 + b_\pi^{SU(2)} m_\pi^3 
+ \ldots
,\end{equation}
where the omitted terms in Eq.~\eqref{eq:MX2} consist of higher powers of the expansion parameter. 
In the above form, 
the non-analytic strange quark mass dependence has been absorbed into the relevant low-energy constants
of the two-flavor chiral expansion of the sigma mass. The convergence of the $\S$ mass is now governed by: 
the chiral expansion, 
$m_\pi^2 / \Lambda_\chi^2$, 
and the heavy $\S$ expansion,
$m_\pi / M_\S^{SU(2)}$. 

An expansion of hyperon observables in powers of 
$\varepsilon_{SU(2)}$  
is very well behaved. 
There are additional expansion parameters which are related to 
 kaon production thresholds.
Clearly for the two-flavor theory to be effective, 
kaon production thresholds cannot be reached. 
Typically loop diagrams in which the baryon strangeness changes have non-negligible mass splittings 
between the external and intermediate-state baryons. 
For example, 
a generic $B' \to K B$ process
is a $\D S = -1$ strangeness changing baryon transition, 
and is characterized by the mass splitting
$\delta_{BB'}$, 
given by
\begin{equation}
\delta_{BB'} = M_{B'} - M_{B}
.\end{equation}
When the splitting exceeds the kaon mass, 
$\delta_{BB'} > m_K$,
the decay is kinematically allowed, 
otherwise the process 
$B' \to K B$ is virtual. 

To deduce the expansion parameter relevant for an 
$SU(2)$ 
description of hyperons, 
we focus on a schematic example, 
and include the 
$SU(3)$ 
splitting, 
$\delta_{BB'}$. 
The introduction of this scale into loop integrals produces a more complicated non-analytic function involving both 
$m_K$ 
and 
$\d_{BB'}$. 
For diagrams of the sunset type, 
a logarithm is generically produced of the form
\begin{equation}
\cL (m_K^2, - \d_{BB'} )
= 
\log 
\left( 
\frac{- \d_{BB'} - \sqrt{\d_{BB'}^2 - m_K^2 + i \epsilon}}{ - \d_{BB'} + \sqrt{\d_{BB'}^2 - m_K^2 + i \epsilon}}
\right)
,\end{equation}
which contains the non-analyticities associated with kaon production. 
Our concern is with the region below threshold. 
In the limit $\delta_{BB'}  \lesssim m_K$, 
the 
$SU(2)$ 
treatment must fail, 
and we must address whether the physical splittings actually put us in this regime. 
Applying the perturbative expansion about the 
$SU(2)$ 
chiral limit for the generic logarithm, 
we make the following observation, 
namely 
terms in the logarithm that are expanded can be written as functions of the form 
\begin{equation} \label{eq:expand}
f \Big( m_K^2 - \delta_{BB'}^2 \Big)
=
f \Big( \frac{1}{2} m_{\eta_s}^2 - \delta_{BB'}^2 \Big)
+
\frac{1}{2} m_\pi^2 \,
f' \Big( \frac{1}{2} m_{\eta_s}^2 - \delta_{BB'}^2 \Big)
+
\frac{1}{8} m_\pi^4 \,
f'' \Big( \frac{1}{2} m_{\eta_s}^2 - \delta_{BB'}^2 \Big)
+ \ldots \, 
.\end{equation}
Thus for the subthreshold case, 
the expansion parameter, 
$\varepsilon_{BB'} $,
is of the form
\begin{equation} \label{eq:discovery}
\varepsilon_{BB'} 
= 
\frac{\frac{1}{2} m_\pi^2}{ \frac{1}{2} m_{\eta_s}^2 -  \delta_{BB'}^2}
.\end{equation}

We can diagnose the convergence properties of the
$SU(2)$
expansion by estimating the size of the expansion parameters governing
the description of kaon thresholds. 
For the 
$\D S = - 1$ 
virtual transitions, 
we have:
$\varepsilon_{N \Sigma^*} = 0.24$, 
$\varepsilon_{\Lambda \Xi^*} = 0.15$, 
$\varepsilon_{\Xi \Omega} = 0.08$, 
$\varepsilon_{\Sigma \Xi^*} = 0.08$,
$\varepsilon_{N \Sigma} = 0.05$, 
$\varepsilon_{\Lambda \Xi} = 0.05$, 
$\varepsilon_{N \Lambda} = 0.04$, 
$\varepsilon_{\D \Sigma^*} = 0.04$,
$\varepsilon_{\Sigma^* \Xi^*} = 0.04$, 
$\varepsilon_{\Xi^* \Omega} = 0.04$, 
and
$\varepsilon_{\Sigma \Xi} = 0.04$,
while for the 
$\D S = 1$ 
virtual transitions, 
the parameters are:
$\varepsilon_{\Lambda \Delta} = 0.04$,  
$\varepsilon_{\Xi \Sigma^*} = 0.04$, 
and
$\varepsilon_{\Sigma \Delta} = 0.04$.
For a majority of the strangeness changing transitions, 
the mass-splittings play little role in the 
$SU(2)$ 
expansion, 
i.e.~$\varepsilon_{BB'} \approx \varepsilon_{SU(2)}$. 
Despite the nearness of thresholds (compared to the kaon mass), 
the expansion parameters in 
$SU(2)$ 
are all better than the generic expansion parameter for 
$SU(3)$, 
$\varepsilon \sim m_\eta / M^{SU(3)} = 0.5$. 

We have estimated the mass of the 
$\eta_s$ 
meson by using the GMOR relation for the neutral kaon mass. 
Allowing this mass to vary 
$10 \%$
shows that one of the transitions listed has a potentially fallible expansion.  
If the mass of the 
$\eta_s$
is 
$10 \%$
smaller,
then the expansion in 
$\varepsilon_{N \Sigma^*}$ 
is ill-fated. 
To further address the convergence of
$SU(2)$
chiral perturbation theory for hyperons, 
we must assess the impact of next-to-leading order corrections, namely
we must find an appropriate expanion parameter for $SU(2)$ beyond leading order. 
This parameter must take into account what is practically done in performing an
$SU(2)$ 
chiral expansion, 
namely re-summing strange quark mass contributions while treating the lighter quark mass dependence perturbatively. 
Beyond leading-order, 
the mass parameters that naturally enter the two-flavor theory are the
$SU(2)$
chiral limit values 
$m_K^{SU(2)}$ and 
$\d_{BB'}^{SU(2)}
\equiv
M_{B'}^{SU(2)} - M_B^{SU(2)}
$.
The 
$SU(2)$
superscripts denote the evaluation of quantities in the limit of vanishing lighter quark masses,
$m_u = m_d = 0$.

To match the 
$SU(3)$ 
loop contributions onto the
$SU(2)$
theory, 
we should only include baryon mass splittings due to the strange quark,
$\d_{BB'}^{SU(2)}$.
Non-analytic contributions associated with the kaon threshold then schematically have the dependence
\begin{equation}
f \Big( m_K^2 -  [\d_{BB'}^{SU(2)} ]^2 \Big)
=
f \Big( m_K^2 - [m_K^{SU(2)}]^2 + [m_K^{SU(2)}]^2  - [\d_{BB'}^{SU(2)} ]^2 \Big)
.\end{equation}
Expanding about the 
$SU(2)$
chiral limit gives an expansion parameter
\begin{eqnarray}
\varepsilon_{BB'} 
&=& 
\frac{m_K^2 - [m_K^{SU(2)}]^2 } {[m_K^{SU(2)}]^2 - [\delta_{BB'}^{SU(2)}]^2}
= \label{eq:NEW}
\frac{ \frac{1}{2} m_\pi^2} {[m_K^{SU(2)}]^2 - [\delta_{BB'}^{SU(2)}]^2} + \ldots \, \, 
.\end{eqnarray} 
Because the expansion parameters are highly sensitive to the value of the denominator, 
it is a good approximation to expand the numerator to leading order. 
In order to deduce the fate of the 
$SU(2)$
expansion, 
we need to estimate the 
$SU(2)$ 
chiral limit values of the mass parameters.  Using lattice data and inputs from phenomenology, we find the values for $\epsilon_{N\Sigma^{*}}$ and
$\epsilon_{\Lambda\Xi^*}$, which are the two worst possible cases for a $SU(2)$ description of hyperons to fail, are given by  $0.31 \le \epsilon_{N\Sigma^{*}}\le 0.37$ and
$ 0.16\le \epsilon_{\Lambda\Xi^*} \le 0.18$, respectively. Having considered the two worst possible baryon transitions, 
although they are larger than what we have estimated in previous section,
we still expect the
$SU(2)$ 
chiral expansion to provide a good description of kaon threshold contributions to hyperon observables.
\vskip-0.5cm
\section{Effects of Kaon Thresholds}
In this section we turn our attention specifically to the hyperon isovector axial charges and hyperon electromagnetic properties
below.

At leading loop order, 
one encounters a variety of diagrams which contribute to thresholds in the evaluation of axial-vector
current matrix elements. Further the evaluation of loop diagrams of these types produces terms proportional to the following 
non-analytic functions
\begin{eqnarray} \label{eq:Ifunc}
\cI (m_K^2,  - \d_{BB'} , - \d_{B''B'})
&=& 
\frac{2}{3} \frac{1}{\d_{BB'} - \d_{B''B'}} 
\Big[ 
\cF(m_K^2, -\d_{BB'})
-
\cF(m_K^2, -\d_{B''B'})
\Big]\nonumber \\
\cJ (m_K^2, - \d_{BB'} )
&=& 
- 2 \d_{BB'} 
( \d_{BB'}^2 - m_K^2)^{1/2} 
\cL \big(m_K^2, - \d_{BB'} \big) 
,
\end{eqnarray}
where ${\cal F}$ is related to ${\cal L}$ introduced earlier by ${\cal F}$ = $-(\delta_{BB'}^2-m_K^2)^{3/2}{\cal L}$. Notice while ${\cal I}$ is associated with an isospin transition and possibly a transition from a spin-half baryon to a spin three-halp baryon or vice versa, ${\cal J}$ appears when evaluating diagrams related to spin-conserving axial current.  
A $SU(2)$ chiral expansion of the non-analytic function ${\cal J }$ up to $m_{\pi}^4$ for $\Sigma$ and $\Sigma^{*}$ with intermediate-state isovector axial transition $N \to N$  are shown in figure 1. Notice the dashed green line in both panel of figure 1 is the non-analytic contributions ${\cal J}$. Further, the red curve is the zeroth-order approximation, 
$\cJ^{(0)}$, 
while the blue curve also includes the first-order correction (which is of order $m_\pi^2$), 
and finally the black curve includes all terms to 
$m_\pi^4$. Figure 1 clearly demonstrates that the virtual kaon loop contributions with intermediate-state isovector axial transition $N \to N$ to the isovector axial charge of $\Sigma$ and $\Sigma^*$ can be captured very nicely by a $SU(2)$ chiral expansion. Similar results are obtained for the non-analytic function ${\cal F}$ as well. These results in turn imply all the virtual kaon contributions to the isovector axial charges of hyperons are well-described by terms in $m_{\pi}^2$.  

\begin{figure}[t]
\begin{center}
\epsfig{file=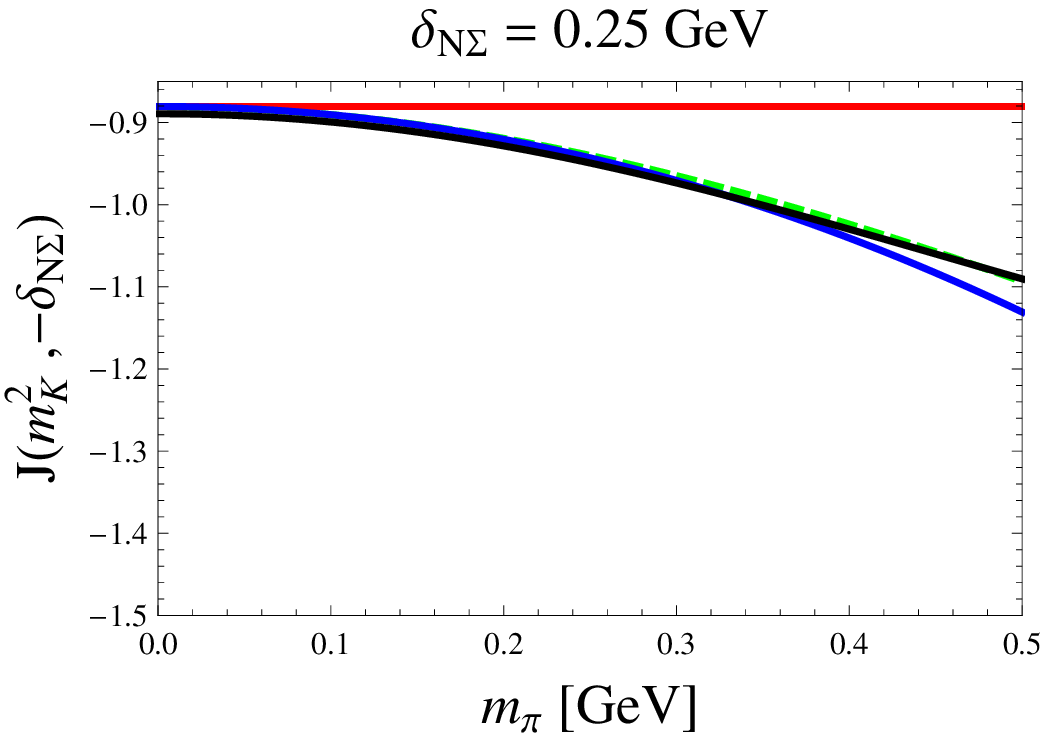,width=2.6in}
$\quad$
\epsfig{file=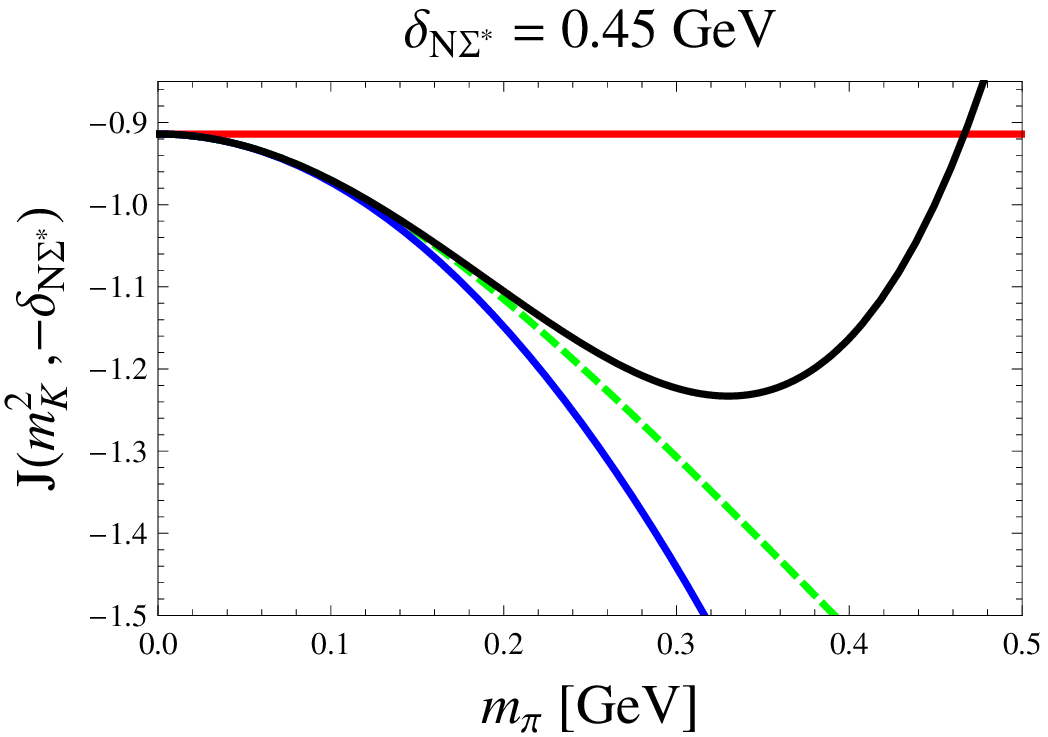,width=2.6in}\vskip-0.4cm
\caption{ 
Virtual threshold contribution from the kaon loop with intermediate-state isovector axial transition
$N \to N$ for $\Sigma$ (left) and
$\Sigma^*$ (right)  baryons.
}
\end{center}
\label{figJ}
\end{figure}

Notice the non-analytic function ${\cal J}$ appears in the virtual loop contributions to the hyperon magnetic moments as well. As a result, the
virtual kaon loop contributions to the hyperon magnetic moments can also be captured nicely with a $SU(2)$ chiral expansion.

Similarly, the non-analytic functions appearing in the loop contributions to hyperon charge radii (quadrupole moments) and quadrupole radii are given by
\begin{eqnarray}
\label{eq:GG}
&&G_{Th} (m_K^2, - \d_{BB'} )
=
\frac{\d_{BB'}}
{\left( \d_{BB'}^2 - m_K^2 \right)^{1/2}} 
\,
\cL (m_K^2, - \d_{BB'} )
,\quad {\text{and}} \nonumber \\
&&{\cal G}_{Th} (m_K^2, - \d_{BB'} )
=
\frac{1}{10}
\left[
\frac{2}
{ \d_{BB'}^2 - m_K^2 } 
-
\frac{\d_{BB'}}{\left( \d_{BB'}^2 - m_K^2 \right)^{3/2}}
\,
\cL (m_K^2, - \d_{BB'} )
\right]
,\end{eqnarray} 
respectively. For $\Sigma^*$, contribution from the $K$-$N$ loop to charge radius (quadrupole moment) and quadrupole radius of
$\Sigma^{*}$ is shown as dashed green line in the left panel and right panel of figure 2. Compared with these curves are the first three approximations in the $SU(2)$ expansion that are analytic in 
$m_\pi^2$.
The red curve is the zeroth-order approximation, 
the blue curve includes the first-order correction proportional to 
$m_\pi^2$,
and finally the black curve includes all terms to 
$m_\pi^4$.  The results in figure 2 shows that the kaon contributions to $\Sigma^*$ charge radius, quadrupole moment and quadrupole radius remain perturbative not far away from physical pion mass. The increased sensitivity in these observables
is due to the threshold singularities in the non-analytic functions, 
Eqs.~\eqref{eq:GG}. 
By contrast, 
kaon contributions to 
the masses, 
axial charges, 
and magnetic moments vanish at the kaon threshold
due to phase-space factors. 
For the case of radii and quadrupole moments, 
the kaon contributions become singular near threshold. Interestingly, a similar situation is observed in \cite{Led10}, namely the appearance of analogous singularities at the pion threshold is indicative of a breakdown in the standard definition of 
moments and radii of resonances.

\begin{figure}[t]
\begin{center}
\epsfig{file=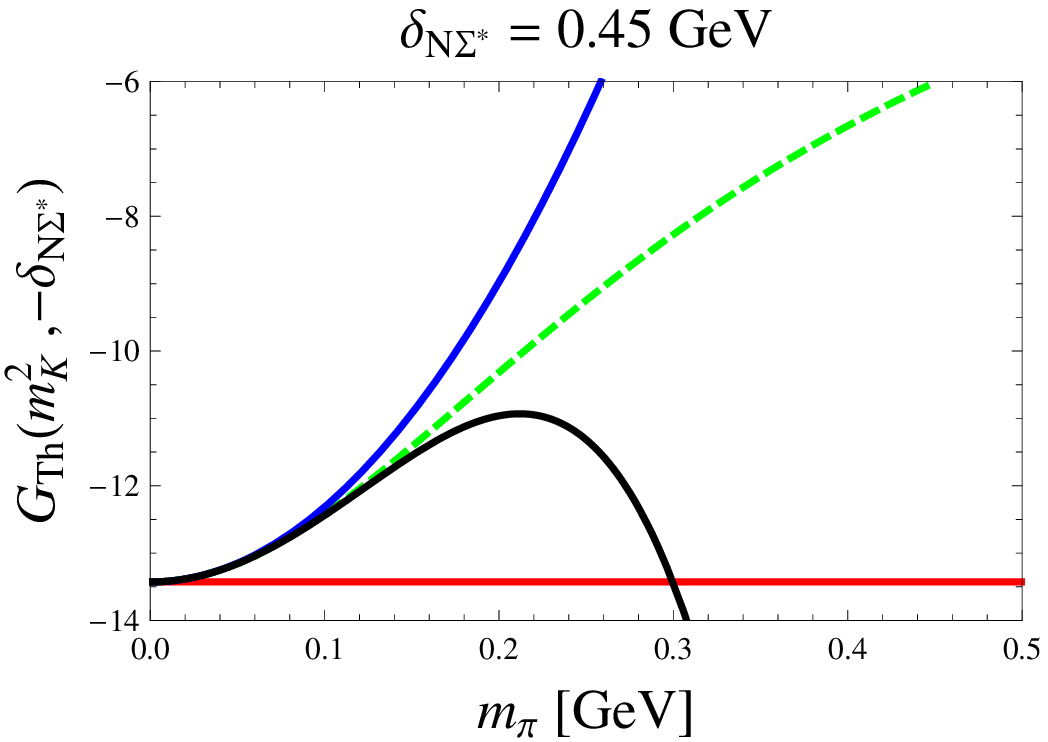,width=2.6in}
$\quad$
\epsfig{file=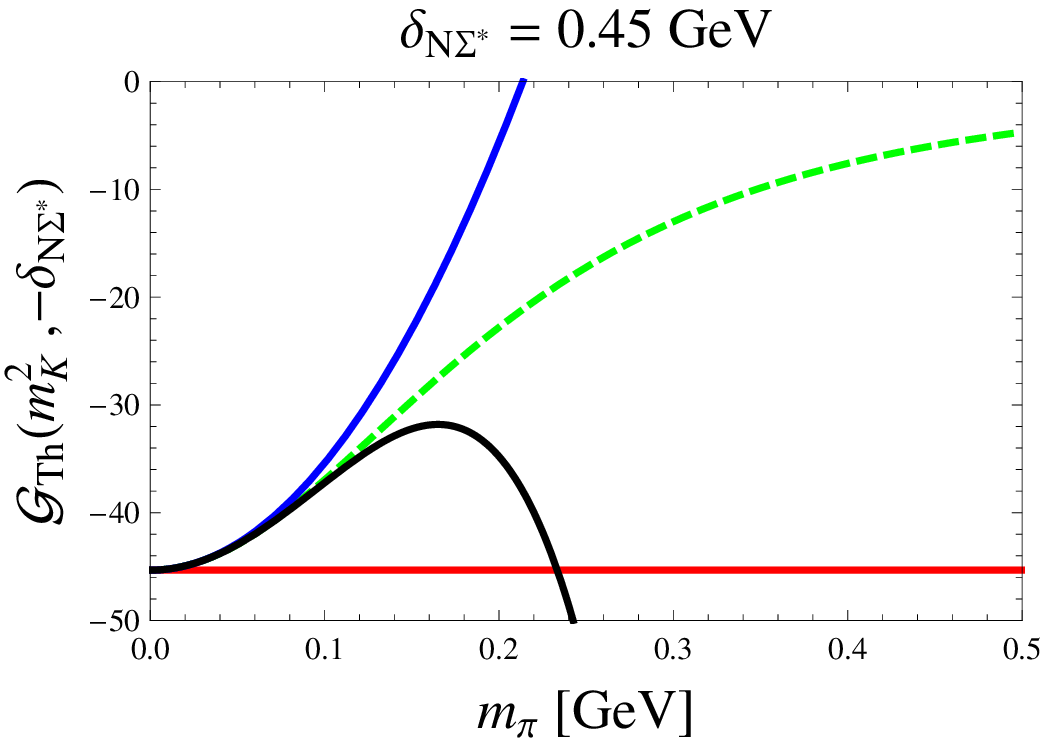,width=2.6in}
\vskip-0.4cm
\caption{\label{f:Gexpand}
Contribution from the
$K$-$N$ 
loop diagram for the charge radii (left) and quadrupole radius (right) of
$\Sigma^*$ baryons.
 }
\end{center}
\label{figGG}
\end{figure}

\section{Isovector Axial Charges of Spin-1/2 Hyperons}
Using phenomenological input and lattice data, we compute the isovector axial charges of spin-half hyperons in the framework of two-flavor chiral perturbation theory for hyperons. Further, we have compared our analytic 
expressions with the lattice results of hyperon axial charges obtained in \cite{Lin07}. Comparing the isovector axial charges obtained, $g_A \sim 1.2$, $g_{\Sigma\Sigma}\sim0.8$ and $g_{\Xi\Xi} \sim 0.2$, suggests better convergence of $\chi PT$ with increasing strangeness quantum number. Our calculations verify this pattern of convergence for the axial charges as can been seen in figure 3.   

\begin{figure}[t]
\label{figaxial}
\begin{center}
\epsfig{file=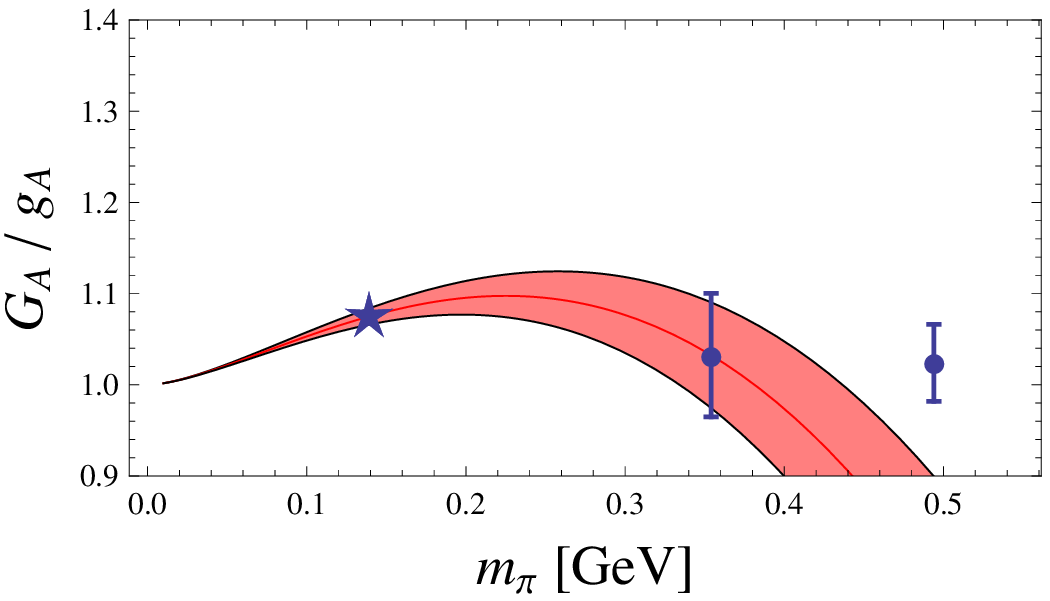,width=2.25in}
$\quad$
\epsfig{file=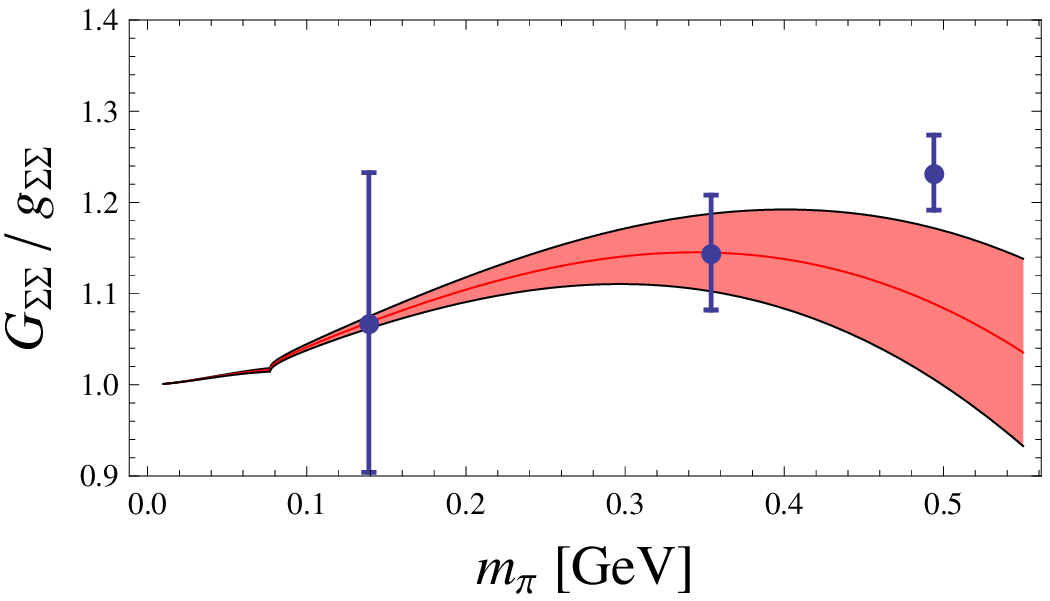,width=2.25in}
$\quad$\vskip-0.15cm
\epsfig{file=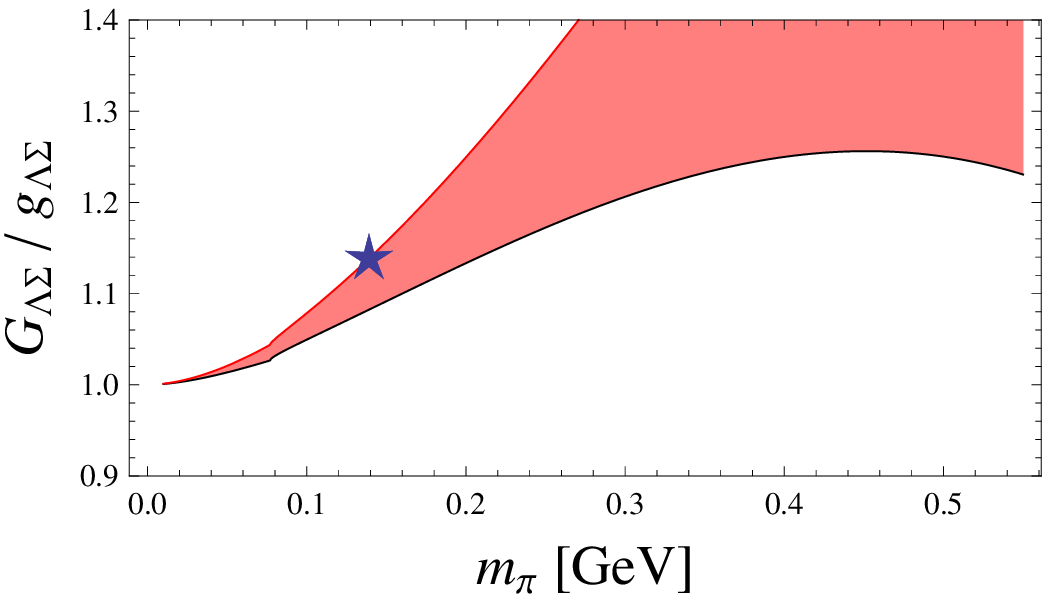,width=2.25in}
$\quad$
\epsfig{file=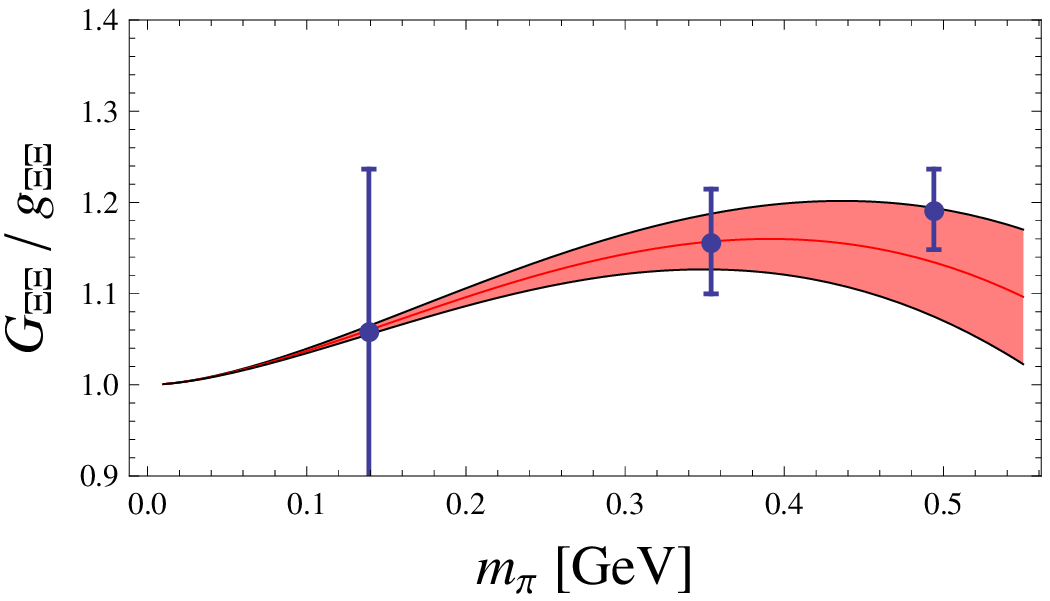,width=2.25in}
\vskip-0.4cm
\caption{\label{axial:chiralcorrection}
Pion mass dependence of isovector axial charges. 
Stars denote physical inputs, 
while the points are lattice QCD results taken from \cite{Lin07},
of which the value at the physical pion mass is obtained from an empirical quark-mass extrapolation,
and the lowest mass data are used to estimate chiral limit couplings and local contributions at NLO. 
 }
\end{center}
\end{figure}

\section{Conclusions}
We use a two-flavor chiral perturbation theory for hyperons to study the axial charges and electromagnetic peroperties of hyperons. In particular we investigate the importance of virtual kaon thresholds on hyperons properties. For the case of experimentally known hyperon magnetic moments, axial charge and electromagnetic charge radii,
we demonstrate that chiral corrections are under reasonable control,
in contrast to the behavior of these observables in the three-flavor chiral expansion. 
The results we obtained are ideal for performing the pion mass extrapolation
of lattice QCD data obtained at the physical strange quark mass. 

\vskip0.15cm

Partial support from DOE and NCTS (North) are acknowledged. The results presented here have already published in
\cite{Tib08,Tib09,Jia09,Jia09_1,Jia10}. F.J. Jiang acknowledges most of his contributions to the results appearing in this proceeding are done at ITP of Bern University and at CTP of MIT.

\end{document}